\documentstyle[twocolumn,epsf]{jpsj}

\def\runtitle{
2D $S=1$ Heisenberg Antiferromagnet at Finite Temperatures
}

\def\runauthor{
Kenji {\sc Harada}, Matthias {\sc Troyer} and Naoki {\sc Kawashima}
}

\title
{
The Two-Dimensional $S=1$ Quantum Heisenberg Antiferromagnet 
at Finite Temperatures
}

\author
{ 
Kenji {\sc Harada}, Matthias {\sc Troyer}$^1$ and Naoki {\sc Kawashima}$^2$
}

\inst
{
Division of Applied Systems Science, Kyoto 
University, Sakyo-ku, Kyoto 606-01\\
$^1$
Institute for Solid State Physics,
University of Tokyo, Roppongi 7-22-1, Tokyo 106\\
$^2$
Department of Physics, Toho University, 
Miyama 2-2-1, Funabashi 274, Japan
}

\recdate
{
December 26, 1997
}

\abst
{
The temperature dependence of the correlation length, susceptibilities
and the magnetic structure factor of the two-dimensional spin-1 square
lattice quantum Heisenberg antiferromagnet are computed by the quantum
Monte Carlo loop algorithm (QMC). In the experimentally relevant temperature
regime the theoretically predicted asymptotic low temperature behavior
is found to be not valid. The QMC results however, agree reasonably well
with the experimental measurements of ${\rm La}_2{\rm NiO}_4$ even without
considering anisotropies in the exchange interactions.
}

\kword
{
Heisenberg model, quantum Monte Carlo
}

\begin{document}
\sloppy \maketitle

Recently, field theoretical predictions~\cite{CHN,HN,CSY} concerning the
correlation length of the square lattice quantum Heisenberg antiferromagnet
(QHA) were directly checked by experimental measurements~\cite{NYH,GBE} and
several quantum Monte Carlo (QMC) simulations~\cite{MD,KLT,BBGW,TK}. While
in the case of spin $S=1/2$ the validity of the predictions seemed
supported by experiments, in the case of $S=1$, experimental measurements
for both ${\rm La}_2 {\rm NiO}_4$~\cite{NYH} and ${\rm K}_2{\rm
  NiF}_4$~\cite{GBE} turned out to be inconsistent with the theoretical
predictions.

The inconsistencies were explained by noting that the theoretical low
temperature expression is valid for the temperatures of the $S=1/2$
experiments, but not valid in the temperature regime of the $S=1$
experiments.  While a number of simulations have been performed for
$S=1/2$, only a high temperature series expansion calculation~\cite{series}
and an effective high temperature theory
(PQSCHA)~\cite{CUCC:1,CUCC:2,CUCC:3} are available for $S=1$ in the
experimentally relevant temperature range. In this paper, we compute the
correlation length and other thermal averages for the $S=1$ QHA on a square
lattice using the quantum Monte Carlo loop algorithm~\cite{LOOP}
generalized to larger spins~\cite{KG:1,KG:1a,KG:2}. This algorithm was
implemented in continuous imaginary-time representation~\cite{BW} to
eliminate the systematic error due to Suzuki-Trotter discretization of path
integrals.

\begin{figure}
\epsfxsize=0.50\textwidth \epsfbox{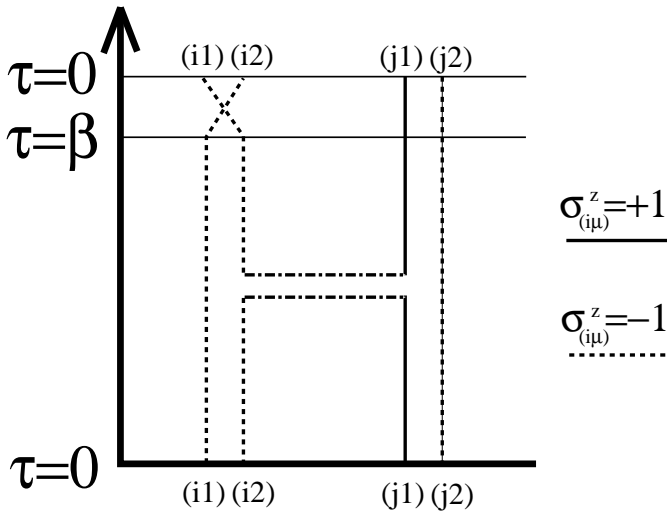}
\caption{
  A cut and two horizontal segments.  The two types of boundary condition
  in temporal direction are also illustrated.  }
\label{fig:cut}
\end{figure}

When implemented in continuous imaginary time, the probability for the
graph assignment in the cluster algorithms for larger spins becomes much
simpler than the original discrete time version ~\cite{KG:1,KG:1a},
although the idea is essentially the same.  As in the case of discrete
time, we first extend the Hilbert space by expressing each spin operator by
a sum of $2S$ Pauli spins:
\begin{equation}
  S_i = \frac12 \sum_{\mu = 1}^{2S} \sigma_{i\mu}.
\end{equation}
We therefore consider $(2SN)$ vertical lines along the imaginary time axis,
each specified by two indices $(i\mu)$, where $N$ is the total number of
original spins.  Since there are unphysical states in the new Hilbert space
in which some of the spins have magnitude less than $S$, we must eliminate
such states by applying projection operators.  Here we use a representation
where $z$-spin components are diagonalized.  Our procedure for the graph
assignment is as follows.  For each pair of neighboring world lines and for
each uninterrupted time interval during which spins on these world lines
are antiparallel, we generate ``cuts'' of worldlines with probability
density $J/2$.  At each cut, we reconnect pairwise the four end points
created by the cut by two horizontal segments (Fig.~\ref{fig:cut}).  The
application of the above-mentioned projection operator for a site is
realized by choosing an appropriate boundary condition in the temporal
direction.  If the spin values $(\sigma_{(i\mu)}^z(\tau))$ are the same at
the four end points, $(i1)$ and $(i2)$ at $\tau = \beta$ and at $\tau = 0$,
we choose a straight connection (connecting $(i1)$ at $\tau = \beta$ to
$(i1)$ at $\tau = 0$) and a cross connection (connecting $(i1)$ at $\tau =
\beta$ to $(i2)$ at $\tau = 0$) with equal probability.  Otherwise, we
choose the unique method for connecting these four points pairwise so that
the spin value at each connection point is continuous.  Thus, we form many
loops which are to be flipped with probability 1/2.  This algorithm turns
out to be more efficient than its discrete version.

Using chiral perturbation theory, Hasenfratz and Niedermayer(HN)~\cite{HN}
obtained the temperature dependence of the correlation length up to
two-loop order for an arbitrary magnitude of spin:
\begin{equation}
  \xi_{HN} = \frac{e}{8} \frac{\hbar c}{2\pi\rho_s} \exp \left(\frac{1}{t}
  \right) \times \left[ 1 - \frac{t}{2} +{\cal O}\left( t^2\right)
  \right],\label{eq:HN}
\end{equation}
where $t\equiv k_{\rm B}T/2\pi\rho_S$.  The dependence on the magnitude of
the spin $S$ is only implicit through the $S$-dependence of the the
spin-stiffness constant $\rho_s$, and the spin wave velocity $c$. For
$S=1/2$, the spin wave theory (SWT)~\cite{SW} values for $\rho_s$ and $c$
were observed to be close to the QMC estimates. As SWT is better for larger
spins, we can confidently use the SWT values for $S=1$ in the following
with $2\pi\rho_s=5.461J$ and $\hbar c=3.067J$. These values agree with the
result of a series expansion around the Ising limit, within statistical
errors smaller than 1\%.

Equation (\ref{eq:HN}) is valid in the renormalized classical regime where
\begin{equation}
  t \ll 1.
\end{equation}
An additional constraint on the temperature comes from the cutoff of the
quantum fluctuations in the effective field theory once the extension in
the imaginary time direction $\beta$ becomes smaller than the lattice
spacing $a/\hbar c$:
\begin{equation}
  t \ll {\hbar c\over 2\pi\rho_s a} \approx {\sqrt{2}\over\pi}{1\over
    S}\left(1+{0.196\over S}
  \right).
\label{eq:trange}
\end{equation}
The approximation in the last term on the right hand side is again the
leading order SWT result.

In the case of $S=1/2$ QHA, it was found by QMC simulations that
eq.~(\ref{eq:HN}) is valid only at very large correlation
lengths~\cite{BBGW,TK} of the order of 100 lattice spacings or larger. In
the experimentally relevant temperature regime the deviations, while
clearly visible in the QMC simulations, are however smaller than the
experimental errors.  Thus, the theory and experiment agree for $S=1/2$.

The large discrepancies observed for $S=1$~\cite{NYH,GBE} are somewhat
counterintuitive, since the theory based on a spin-wave picture should be
better for larger spins.  However, this is not very surprising since
eq.~(\ref{eq:HN}) reflects the quantum nature of the system and therefore
eq.~(\ref{eq:HN}) applies only if eq.~(\ref{eq:trange}) is satisfied.  The
region determined by eq.~(\ref{eq:trange}) should be smaller for systems
closer to the classical system.  Therefore, since larger spins are more
classical as noted previously in refs.~\citen{CHN,BBGW,TK,series}, the
validity of eq.~(\ref{eq:HN}) is restricted to even larger correlation
lengths than for $S=1/2$, much larger than accessible in experiments.

As the analytic low temperature form is not valid in the experimentally
accessible temperature regime, we performed QMC simulations over the
temperature range $t\ge 0.18$ ($k_{\rm B}T \ge J$) in order to compare the
experimental data~\cite{NYH} with the $S=1$ QHA.  The second moment
correlation length $\xi_L$ on a finite system of size $L$ was determined
from the static structure factor $S_L({\mib q})$ in the vicinity of ${\mib
  Q}=(\pi,\pi)$~\cite{BK}. The $\xi_L$ is calculated as follows.
\begin{equation}
  \label{xi-def}
  \xi^{-2}_L \equiv f(2\pi/L,0)+f(0,2\pi/L)-f(2\pi/L,2\pi/L),
\end{equation}
where
\begin{equation}
  \label{f-def}
  f({\mib q}) \equiv 4 \sin^2\left(\frac{{\mib q}}{2}\right)
  \left(1-\frac{S({\mib Q}+{\mib q})}{S({\mib Q})}\right)^{-1}.
\end{equation}
This estimator for the correlation length (of a finite system) should be
correct up to the fourth order in $2\pi/L$.  We use improved estimators to
reduce statical errors.  For a given temperature $T$, when the lattice size
$L$ is sufficiently large, $\xi_L$ converges to a size-independent value
$\xi(T)$, which we regard as the infinite-size limit.  We find that the
convergence is achieved to the accuracy determined by the present
statistical error when the condition $L \ge 7 \xi_L$ is satisfied.  All the
following results are obtained under this condition. For each simulation we
have performed $10^6$ sweeps, after $10^4$ thermalization sweeps.

\begin{figure}
\epsfxsize=0.5\textwidth\epsfbox{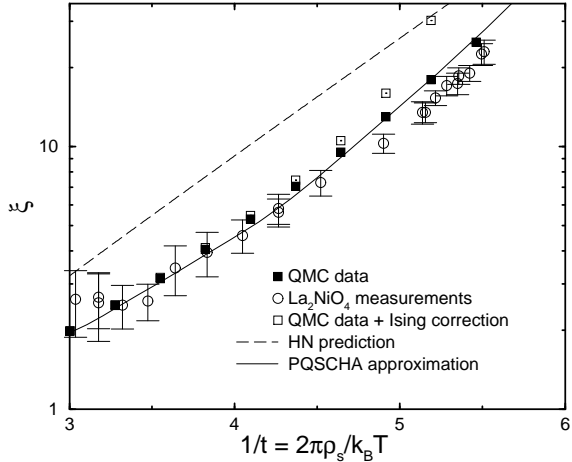}
  \caption{The correlation length $\xi$ as a function of
  $t=k_{\rm B}T/2\pi\rho_s$.  Filled squares represent the results of our
  simulation, the open circles are experimental
  measurements\protect~\cite{NYH}. While the experimental measurements
  agree roughly with our QMC results, they are incompatible with the
  QMC results corrected for a small Ising anisotropy (open squares).
  Compared to analytic calculations, we find that in this temperature range
  the low-temperature predictions of eq.~(\protect\ref{eq:HN}) (dashed line) are
  not valid. The PQSCHA approximation\protect~\cite{CUCC:1,CUCC:2,CUCC:3}, however, agrees
  well with both our data and the experimental measurements.}
\label{fig:xi-beta}
\end{figure}

A selection of our results for the correlation length $\xi$, the staggered
structure factor $S(\pi,\pi)$ and the uniform susceptibility $\chi$ are
summarized in Table~\ref{tab:xi-beta}. In Fig.~\ref{fig:xi-beta}, we plot
our QMC results for the correlation length together with the experimental
data~\cite{NYH} and theoretical predictions based on eq.~(\ref{eq:HN}). Our
present estimates are in rough agreement with experimental measurements.
Greven {\it et al.}~\cite{GBE} and Nakajima {\it et al.}~\cite{NYH} have
additionally proposed to include the effects of a small Ising anisotropy
using a mean-field type correction to the theoretical isotropic
results. This correction (eq.~(7) of ref.~\citen{NYH}) however makes the
agreement worse, as can be seen from Fig.~\ref{fig:xi-beta}.

Compared to theoretical predictions we find that the data deviate strongly
from the low temperature formula of eq.~(\ref{eq:HN}).  The effective
PQSCHA approximation~\cite{CUCC:1,CUCC:2,CUCC:3}, which is an effective
high temperature theory, however agrees well with the QMC results much
better than for $S=1/2$~\cite{KLT,CUCC:1,CUCC:2,CUCC:3}.

\begin{table}
\caption{Correlation length $\xi$, magnetic structure factor $S(\pi,\pi)$
and uniform susceptibility $\chi$
as a function of temperature $t= k_{\rm B}T/2\pi\rho_s$.}
\begin{tabular}{c c c r@{}c@{}l r@{}c@{}l r@{}c@{}l}
\hline
\hline
$J/k_{\rm B} T$ & $t$ & $L$ & \multicolumn{3}{c}{$\xi$} & \multicolumn{3}{c}{$S(\pi,\pi)$} & \multicolumn{3}{c}{$\chi$}\\
\hline
    0.10 & 1.83  &   20  &  0&.&298(5)  &  0&.&8915(1) & 0&.&051579(6)\\
    0.15 & 1.22  &   20  &  0&.&396(5)  &  1&.&0489(2) & 0&.&068503(9)\\
    0.20 & 0.92  &   20  &  0&.&497(5)  &  1&.&2481(3) & 0&.&08118(1)\\
    0.25 & 0.73  &   20  &  0&.&610(5)  &  1&.&5029(5) & 0&.&09038(1)\\
    0.30 & 0.61  &   20  &  0&.&739(5)  &  1&.&8342(7) & 0&.&09687(2)\\
    0.35 & 0.52  &   20  &  0&.&894(6)  &  2&.&272(1) & 0&.&10108(2)\\
    0.40 & 0.46  &   20  &  1&.&079(6)  &  2&.&854(2) & 0&.&10343(2)\\
    0.45 & 0.41  &   20  &  1&.&307(6)  &  3&.&648(2) & 0&.&10435(3)\\
    0.50 & 0.37  &   30  &  1&.&604(3)  &  4&.&776(1) & 0&.&104000(8)\\
    0.55 & 0.33  &   30  &  1&.&982(8)  &  6&.&392(4) & 0&.&10275(2)\\
    0.60 & 0.31  &   40  &  2&.&482(5)  &  8&.&795(2) & 0&.&100756(8)\\
    0.65 & 0.28  &   40  &  3&.&14(1)  &  12&.&46(1) & 0&.&09829(3)\\
    0.70 & 0.26  &   50  &  4&.&06(1)  &  18&.&25(2) & 0&.&09560(2)\\
    0.75 & 0.24  &   50  &  5&.&30(2)  &  27&.&60(3) & 0&.&09275(3)\\
    0.80 & 0.23  &   60  &  7&.&06(2)  &  43&.&30(5) & 0&.&09000(3)\\
    0.85 & 0.22  &   80  &  9&.&52(3)  &  70&.&03(8) & 0&.&08746(3)\\
    0.90 & 0.20  &  120  &  12&.&98(4)  &  116&.&4(1) & 0&.&08516(3)\\
    0.95 & 0.19  &  140  &  17&.&97(5)  &  198&.&8(3) & 0&.&08309(3)\\
    1.00 & 0.18  &  200  &  24&.&94(7)  &  344&.&0(4) & 0&.&08133(2)\\
\hline
\hline
\end{tabular}
\label{tab:xi-beta}
\end{table}

Another violation of the theoretical predictions~\cite{CHN,CSY} was
observed in the peak value of the staggered structure factor $S(\pi,\pi)$,
which according to the theory~\cite{CHN,CSY,CHU} should scale as
\begin{equation}
  \label{eq:sq-1}
  S(\pi,\pi) = A2\pi M^2\xi^2 t^2(1+Ct).
\end{equation}
Here $M$ is the staggered magnetization of the ground state and $A$ and $C$
are universal constants. For spin-1/2 the leading $t^2$ form was confirmed
by high temperature series~\cite{series}. Recent QMC data~\cite{TK} at
lower temperatures fit the above form very well in the temperature range
$t<3$ with $A\approx4.0(1)$ and $C\approx0.5(1)$.  However, experiments for
both $S=1/2$ and $S=1$ were better described by an empirical
law~\cite{NYH,GBE}
\begin{equation}
  \label{eq:sq-2}
  S(\pi,\pi) \propto \xi^2
\end{equation}
over the same temperature range.

We applied a $\chi^2$ analysis to check the consistency of the spin-1 data.
Good fits were obtained for $t<0.28$, with $\chi^2\sim1$ when we allow $C$
to vary, and $\chi^2\sim2$ with a fixed $C=0.5$.  The universal constant $A$
was determined to be $A=4.1(1)$ and $A=4.5(1)$ , similar to the values
obtained for $S=1/2$. The discrepancies between the fits are non-universal
effects caused by the rather high temperatures.

\begin{figure}
\epsfxsize=0.5\textwidth \epsfbox{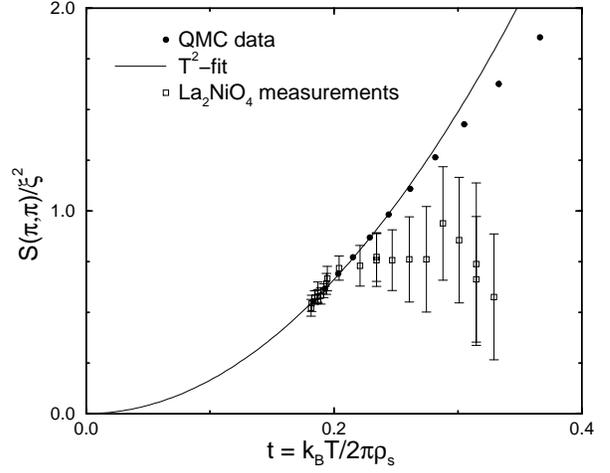}
\caption{The ratio of structure factor peak value and square of the correlation
length $S(\pi,\pi)/\xi^2$ as a function of temperature. Solid circles
represent the QMC measurements and open squares the experimental
measurements. They agree at low temperatures, but differ at higher
temperatures. The solid line is a fit of the QMC data to the
theoretical low-temperature prediction.}
\label{fig:ratio}
\end{figure}

Comparison of our data with the experiments is shown in
Fig.~\ref{fig:ratio} and we can see that for low temperatures the
experimental data are consistent with the QMC results and
eq.~(\ref{eq:sq-1}). The discrepancies that lead refs.~\citen{NYH} and
\citen{GBE} to predict eq.~(\ref{eq:sq-2}) occur at higher temperatures
where the experimental data has large error bars. In view of the precision
of our QMC results and the large errors of the experimental results we
suspect that, contrary to the suggestion of refs.~\citen{NYH} and
\citen{GBE}, the deviations from eq.~(\ref{eq:sq-1}) are due to
uncertainties in the experimental measurements.

\begin{figure}
\epsfxsize=0.5\textwidth \epsfbox{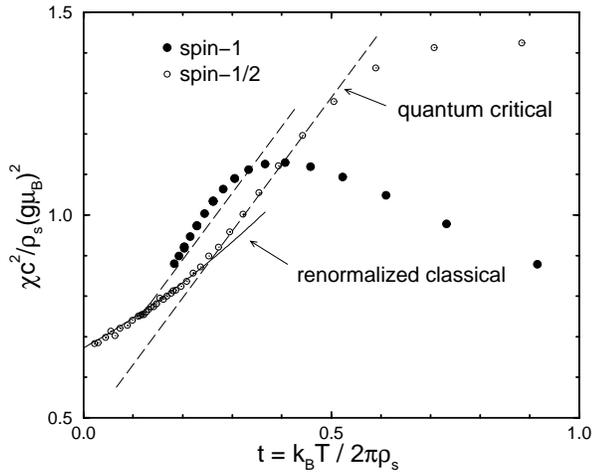}
\caption{The uniform susceptibility $\chi$ as a function of temperature $t$
for both spin $S=1$ and spin $S=1/2$ (taken from
ref.~\protect\citen{TK}).  The solid line is the predicted low
temperature form of the renormalized classical regime. The dashed
lines have the universal slope expected for the quantum critical
regime. As expected, contrary to $S=1/2$, no extended quantum critical
regime exists for $S=1$. However the slope is close to the quantum
critical value as non-universal corrections are still small for $S=1$.}
\label{fig:chiu}
\end{figure}

Finally, we present in Fig.~\ref{fig:chiu} the uniform susceptibility for
$S=1/2$ together with previously published $S=1/2$ results.  First we note
that, as expected, the asymptotic low temperature behavior of the
renormalized classical regime
\begin{equation}
  \chi={2\rho_s\over3}\left({g\mu_B\over\hbar c}\right)^2(1+t+t^2)
\end{equation}
sets in at lower temperatures for $S=1$ than for $S=1/2$.

It was discovered that for $S=1/2$, the uniform susceptibility is the only
quantity for which a clear crossover to quantum critical
behavior~\cite{CSY} can be observed at intermediate temperatures
$t\sim1/3$.  The uniform susceptibility in the quantum critical regime is
\begin{equation}
  \chi=\chi_{\perp}+B T \left({g\mu_B\over\hbar c}\right)^2,
\end{equation}
with a universal slope $B\approx0.26(1)$~\cite{CSY,TI:1,TI:2}.  For $S=1$
however, as discussed above, non-universal corrections become important at
lower temperatures. No quantum critical behavior was thus expected for
$S=1.$ We can see in Fig.~\ref{fig:chiu} that the uniform susceptibility
for $S=1$ deviates from its universal quantum critical behavior at
intermediate temperatures. Its slope is however still surprisingly close to
the quantum critical one, indicating that the non-universal corrections are
still not very large for $S=1$.

In summary, we have simulated the spin $S=1$ quantum Heisenberg
antiferromagnet on a square lattice in the experimentally relevant
temperature regime $k_{\rm B}T\sim J$. We find a better agreement between
the Heisenberg model and experimental data than is expected from the low
temperature theory.  However, in view of the existing small discrepancies
it may be necessary to perform simulations on a model with small
anisotropies in the exchange interactions, and to critically check the data
analysis of the experiments.

\section*{Acknowledgments}
We are grateful to A.~Chubukuv, S.~Sachdev, S.~Chakravarty, Y.~Endoh
and K.~Ueda for helpful discussions. We wish to thank A.~Cuccoli,
V.~Tognetti, R.~Vaia and P.~Verrucchi for providing the results of
their PQSCHA theory and to K.~Nakajima for the experimental data on
${\rm La}_2{\rm NiO}_4$. M.T.\ acknowledges the Aspen Center for
Physics which enabled fruitful discussions that were important for
this work. The calculations were performed on the Hitachi SR2201
massively parallel computer at the computer center of the University
of Tokyo.
N.K.'s work is supported by a Grant-in-Aid
for science research (No.09740320)
from the Ministry of Education, Science and Culture.

\end{document}